\documentclass[article,aps,prd,twocolumn]{revtex4}

\usepackage{graphicx}
\usepackage{amsmath,amssymb,amsfonts}

\begin{document}

\title{Diagnosing the Quark-Gluon Plasma}
\author{Berndt M\"uller}
\address{Department of Physics, Duke University, Durham, NC 27708-3025, USA}
 
\begin{abstract}
Brief review of the hadronic probes that are used to diagnose the quark-gluon plasma produced in relativistic heavy ion collisions and interrogate its properties. Emphasis is placed on probes that have significantly impacted our understanding of the nature of the quark-gluon plasma and confirmed its formation. This report follows closely a prize lecture with same title and contents presented on-line on April 19, 2021 at the Spring Meeting of the American Physical Society.
\end{abstract}

\maketitle 

\section{Introduction}
\label{sec:diag-intro}

Hadronic probes \index{hadronic probes} were recognized early as promising probes for the discovery and study of the quark-gluon plasma. This brief review outlines why they are critical probes, why the effort to validate them met many challenges, and how these probes ultimately proved to be pivotal tools in the recognition of the formation of a new phase of matter containing mobile quarks.

Before delving into the subject itself, it is worth recalling how the need for diagnostics of the quark-gluon plasma (QGP) arose. In 1965, the discovery of cosmic background radiation \index{cosmic background radiation} showed that the universe was exceedingly hot in its infancy. This raised the question of what state of matter filled the universe at these extreme temperatures. Also in 1965 coincidentally, Hagedorn showed that the highest temperature \index{Hagedorn temperature} at which matter consisting of hadrons can exist is approximately 160 MeV, raising the question of what happens beyond that limit. In the late 1970s it became clear that the most likely candidate for matter at these higher temperatures would contain mobile quarks and gluons, in other words, that it must be a QGP.

The nascent scientific community which was interested in this topic at that time, faced two predominant challenges. The first one was: How could a QGP be produced and studied in the laboratory? The other question was: How could such a QGP, once formed, be detected and investigated? 

Nuclear collisions were seen as the only viable path to the creation sufficiently high temperatures to produce a QGP in the laboratory, but it was unclear at the time how much energy exactly was needed. 

European physicists proceeded to inject heavy ions into an existing facility, the CERN Super Proton Synchrotron (SPS) complex for fixed target experiments, while American scientists embarked on building a dedicated facility, the Relativistic Heavy Ion Collider in parallel to exploiting the BNL Alternate Gradient Synchrotron (AGS) for fixed target experiments at lower energies. Theorists at the same time began speculating as to what might be useful QGP signatures. Many, if not most, nuclear physicists were greatly skeptical that any of the proposed signatures would be compelling. And so, as the experiments at the SPS progressed and RHIC was under construction, the Nuclear Science Advisory Committee was asked in 1995 to provide a list of "smoking gun" observables that could tell us whether a QGP was formed and how it behaved.

\begin{figure}[htb]
	\centering
	\includegraphics[width=\linewidth]{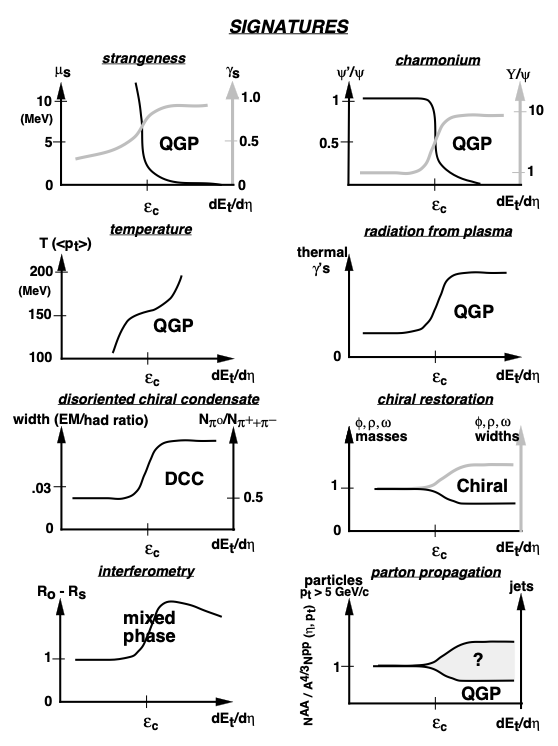}
	\caption{The quark-gluon plasma signatures identified by NSAC in its 1996 Long Range Plan for Nuclear Science \cite{nsac:1996} 
	as primary targets for the experiments at the Relativistic Heavy Ion Collider (from Harris and M\"uller \cite{Harris:1996zx}).}
	\label{fig1:smokingguns}
\end{figure}

Figure~\ref{fig1:smokingguns} shows the smoking guns that NSAC came up with~\cite{nsac:1996,Harris:1996zx}, While many of the signatures studied today are already in this list, there were some glaring omissions. Example of missing signatures are valence quark recombination, anisotropic collective flow, and critical fluctuations. In other words, the community started exploring the problem with a somewhat incomplete set of observables that were considered important.

\section{Diagnostics focused on quarks}
\label{sec:diag-quarks}

Let's now move forward to what we know now and what we have found over the past 20 years. I start with observables that concern the bulk of the QGP and quarks inside the QGP \index{QGP diagnostics} \index{QGP signatures}:
\begin{itemize}
\item
Quark flavor equilibration: that strange quark pairs are produced as abundantly as up and down quark pairs. 
\item
Hadronization at the phase boundary: that quarks recombine into hadrons once the QGP cools down to the Hagedorn temperature and hadrons form. 
\item
Fluctuations of and correlations among conserved quantum numbers that are carried by quarks.
\end{itemize}

\subsection{Strangeness}
\label{subsec:strangeness}

So let's begin with strangeness. \index{strangeness production} The original idea was mainly focused on QGPs produced at the SPS and, possibly, the AGS, because there one expected an abundance of quarks over antiquarks as nuclear matter would be partially stopped. Rafelski pointed out in 1980 that this leads to a suppression of anti-up and anti-down quarks, where anti-strange quarks would not be suppressed and so would be significantly enhanced compared to anti-up and anti-down quarks~\cite{Rafelski:1980rk,Rafelski:1982ii}. In a publication~\cite{Rafelski:1982bi} he stated the following \index{multi-strange baryons}: ``Thus, we almost always have more $\overline{s}$ than $\overline{u}$ and $\overline{d}$ quarks and, in many cases of interest, $\overline{s}/\overline{q} \approx 5$.... When the quark matter dissociates into hadrons, some of the numerous $\overline{s}$ quarks may, instead of being bound into a ($q\overline{s}$) kaon, enter into a ($\overline{q}\overline{q}\overline{s}$) or ($\overline{q}\overline{s}\overline{s}$) anti-baryons, in particular a $\overline{\Lambda}$, the $\overline{\Sigma}$ or $\overline{\Xi}$.'' This was a bold prediction at the time (and it did not even mention the $\Omega$, $\overline{\Omega}$ hyperons!) but it has, over the past two-and-a-half decades, been confirmed by numerous experiments at various facilities that study QGP over two orders of magnitude in collision energy.

\begin{figure}[tb]
	\centering
	\includegraphics[width=0.6\linewidth]{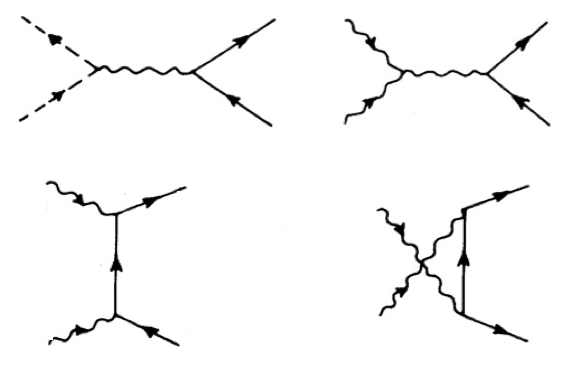}
	\caption{Lowest-order QCD diagrams describing production of strange quark pairs from both gluons and light quark pairs (adopted from Ref.~\cite{Rafelski:1982pu}).}
\label{fig2:QCD-diagrams}
\end{figure}

\begin{figure}[tb]
	\centering
	\includegraphics[width=0.65\linewidth]{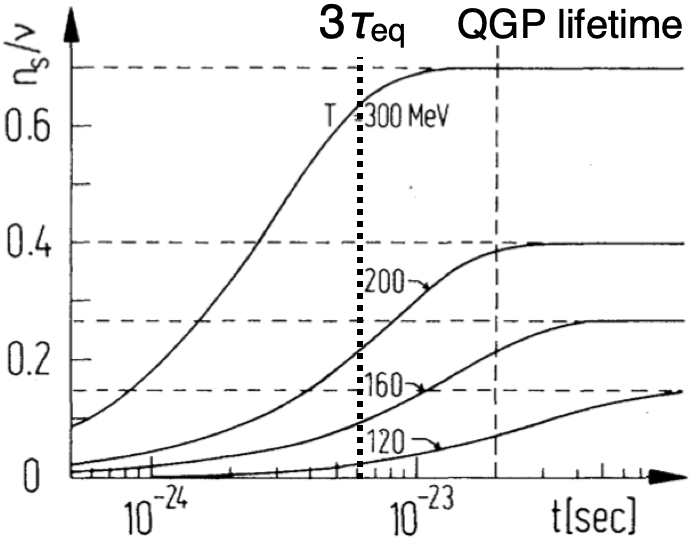}
	\caption{Approach to equilibrium of the strange quark abundance in the QGP for a fixed temperature (adopted from Ref.~\cite{Rafelski:1982pu}).}	
	\label{fig3:s-quarkproduction}
\end{figure}

When this idea was proposed, the following question arose immediately: \index{equilibration, strangeness} \index{equilibration, chemical} Does strangeness actually reach chemical equilibrium during the short lifetime of a QGP? The calculation that we did at the time~\cite{Rafelski:1982pu} was to use the lowest-order QCD diagrams (see  Fig.~\ref{fig2:QCD-diagrams}) to calculate how quickly strange quarks and anti-quarks would equilibrate if the QGP started out hot, but without an abundance of strange quarks in them as they are not present in the colliding nuclei. What we found was that, within the limits of the calculation, strangeness would be equilibrated well within the lifetime of the fireball (see Fig.~\ref{fig3:s-quarkproduction}). In order to obtain this result it was essential to include as seen in Fig.~\ref{fig2:QCD-diagrams} the process in which two gluons fuse forming a $s\overline{s}$ quark pair.

Peter Koch, who was a graduate student at that time, did a very detailed calculation and showed that this equilibrated strange quark population survives final-state interactions once hadrons are formed~\cite{Koch:1986hf,Koch:1986ud}. Also, as it became clear that gluons and quarks acquire thermal masses inside the QGP, the question arose, what happens if gluons become less abundant by virtue of their thermal mass. Peter L\'evai and Tamas Bir\'o pointed out that actually in this case, there is a one-to-two process in which heavy gluons decay into quark anti-quark pairs which adds to the production rates so that equilibration still occurs during the lifetime of the QGP~\cite{Biro:1990vj}. In modern versions of quark-gluon transport theory, this mechanism is included as an off-shell splitting process.

\begin{figure}[tb]
\centering
	\includegraphics[width=0.65\linewidth]{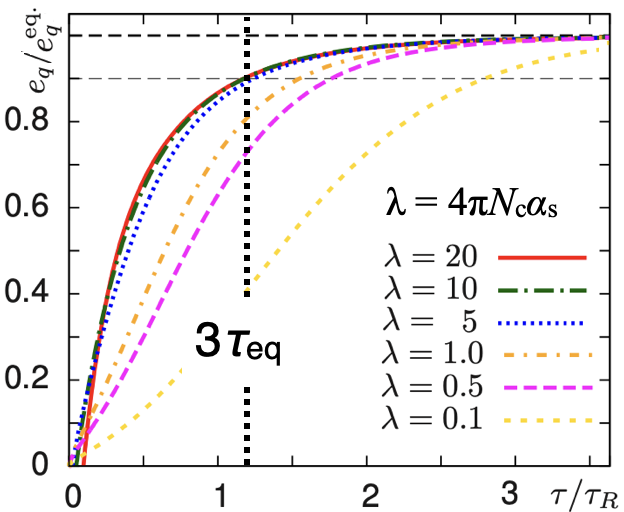}
	\caption{Chemical equlibration rate of ``light'' quarks in a QGP that is initially dominated by gluons. The calculation assumes that the Higgs mass of the quarks is small compared with their thermal mass (adopted from Ref.~\cite{Kurkela:2018oqw}).  $\lambda=4\pi\alpha_s N_c$ denotes the 't Hooft coupling, with $\lambda\approx 10$ at experimentally relevant temperatures.}
	\label{fig4:lightquarkequilibration}
\end{figure}

The last few years have seen a lot of progress in understanding how to deal with the infrared divergences in a thermal QGP. Building on this progress, Aleksi Kurkela and Aleksas Mazeliauskas recently performed a calculation, with all the bells and whistles of today's technology, of how light quarks are chemically equilibrated~\cite{Kurkela:2018xxd,Kurkela:2018oqw}. \index{equilibration, flavor} What they found is that the original results hold up (see Fig.~\ref{fig4:lightquarkequilibration}); the relevant 't Hooft coupling at  the temperature of a few hundred MeV is $\lambda\approx 10$. In other words, strange quarks reach chemical equilibrium quickly in the QGP, and thus strange hadrons should be produced in equilibrium abundances.

\begin{figure}[htb]
	\centering
	\includegraphics[height=3truein]{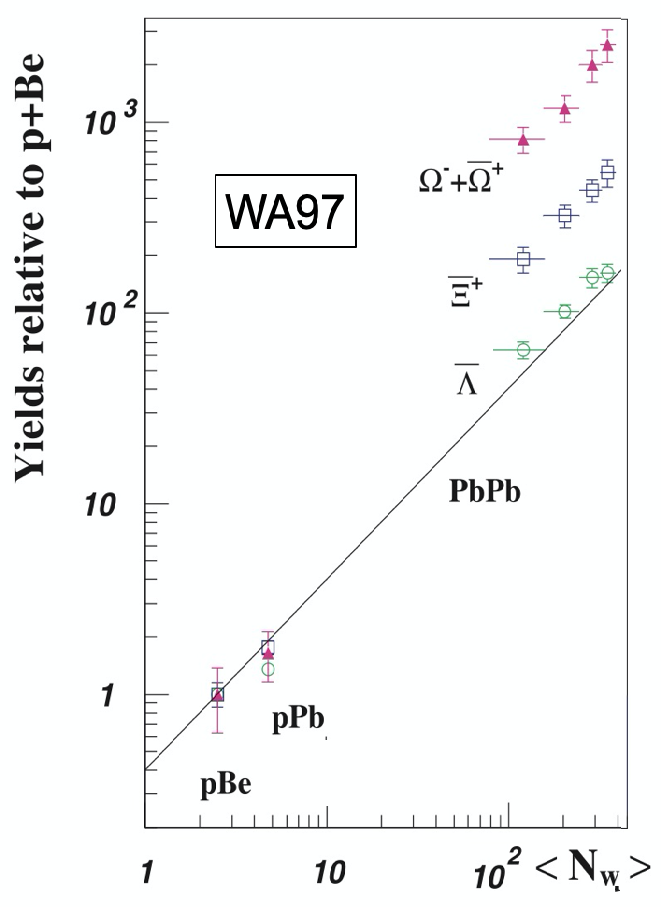}
	\hspace{0.04\linewidth}
	\includegraphics[height=3truein]{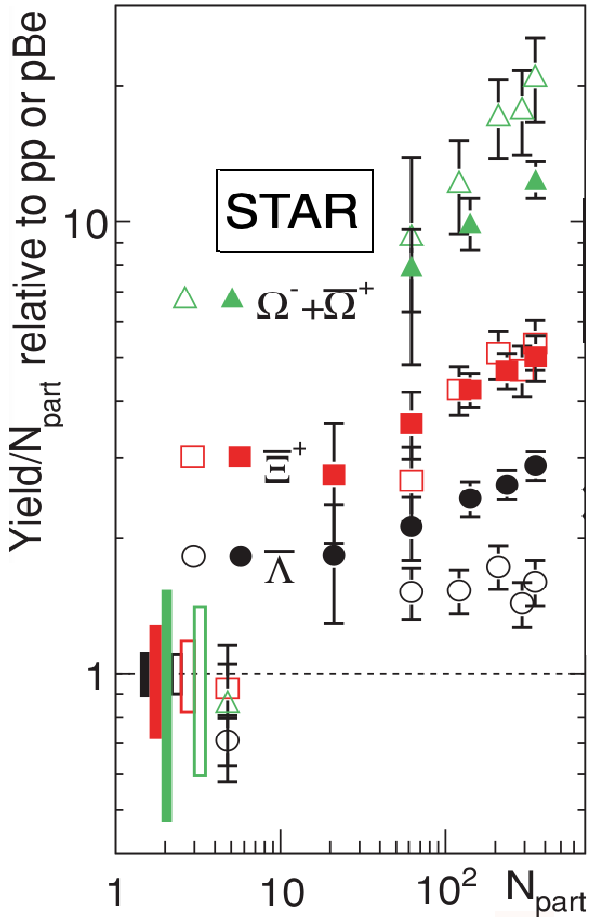}
	\caption{Experimental results showing evidence for enhancement of multi-strange baryons in heavy ion collisions.  Left panel: Pb+Pb collisions at SPS (adopted from WA97 CERN SPS collaboration Ref.~\cite{Antinori:2001yi});  right panel: The solid markers  for Au-Au at $\sqrt{s_{NN}}=200$\,GeV results from STAR at RHIC; the open markers  for Pb+Pb CEREN SPS WA97 results for $|y|<0.5$, $\sqrt{s_{NN}}=17.3$\,GeV   
	(adopted from STAR collaboration Ref.~\cite{Abelev:2007xp}).}	
	\label{fig5:WA97-STAR}
\end{figure}

The following figures highlight a few of the experimental results. The left panel of Fig.~\ref{fig5:WA97-STAR} shows the original results from the SPS, by the WA97 experiment~\cite{Antinori:2001yi}, \index{WA97 experiment} \index{multi-strange baryons} which demonstrated that the production of multi-strange baryons is greatly enhanced -- that of the $\Omega$ and $\overline{\Omega}$ by about a factor of 20 -- compared with what was expected from proton-proton collisions. These results have been also obtained at RHIC by the STAR experiment~\cite{Abelev:2007xp} \index{STAR experiment} (see central panel of Fig.~\ref{fig5:WA97-STAR}). Thus, the enhancement has been confirmed.

\begin{figure}[tb]
	\centering
	\includegraphics[height=4.5truein]{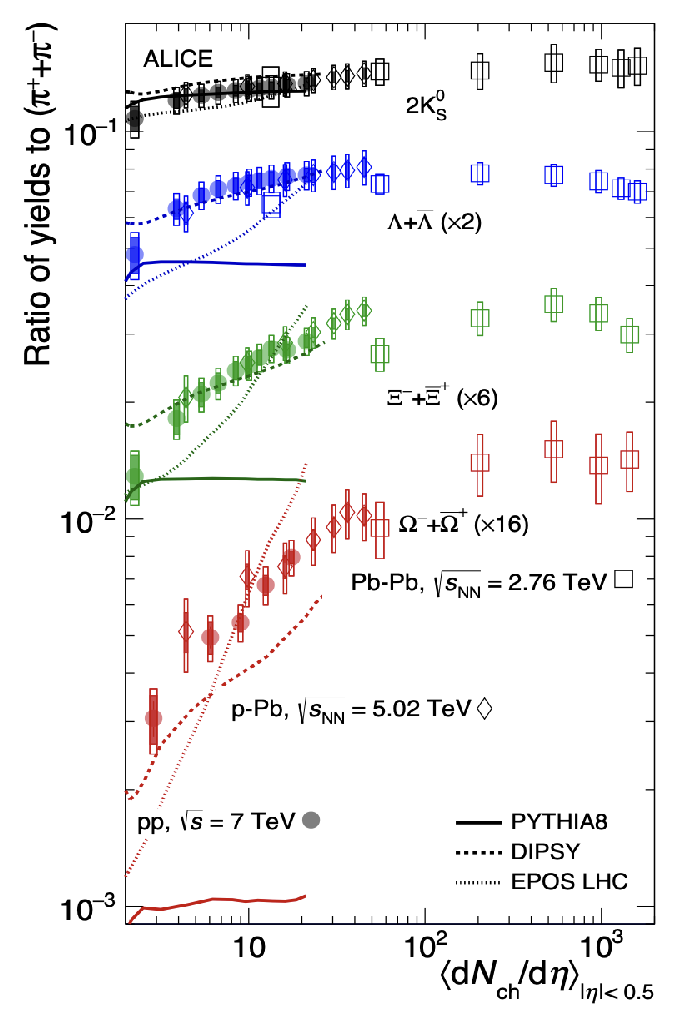}
	\caption{Experimental results showing evidence for enhancement of multi-strange baryons in p+p, p+Pb and Pb+Pb collisions  at LHC as function of the multiplicity of the final state, which is considered as a proxy for the size of the fireball: 
	(ALICE, adopted from Ref.~\cite{ALICE:2017jyt}).}
	\label{fig6:ALICE}
\end{figure}

The ALICE experiment \index{ALICE experiment} \index{equilibration, strangeness} at LHC has recently added to this picture, because they have been able to actually study the enhancement of the strange quark production as a function of the multiplicity of the final state~\cite{ALICE:2017jyt}. As Fig.~\ref{fig6:ALICE} shows, strangeness equilibration develops gradually with increasing fireball size and lifetime, going from increasingly violent proton-proton collisions, to proton-lead collisions and, finally, lead-lead collisions. There are two separate effects that contribute here: one is that in a small fireball net strangeness conservation reduces the overall abundance, called canonical equilibrium, and in a short-lived fireball there is not enough time to reach chemical equilibrium. An interesting thing to do would be to use these experimental data to determine the strange quark equilibration time from the data. This is clearly within reach.

\subsection{Hadronization}
\label{subsec:hadronization}

\begin{figure}[tb]
	\centering
	\includegraphics[width=0.8\linewidth]{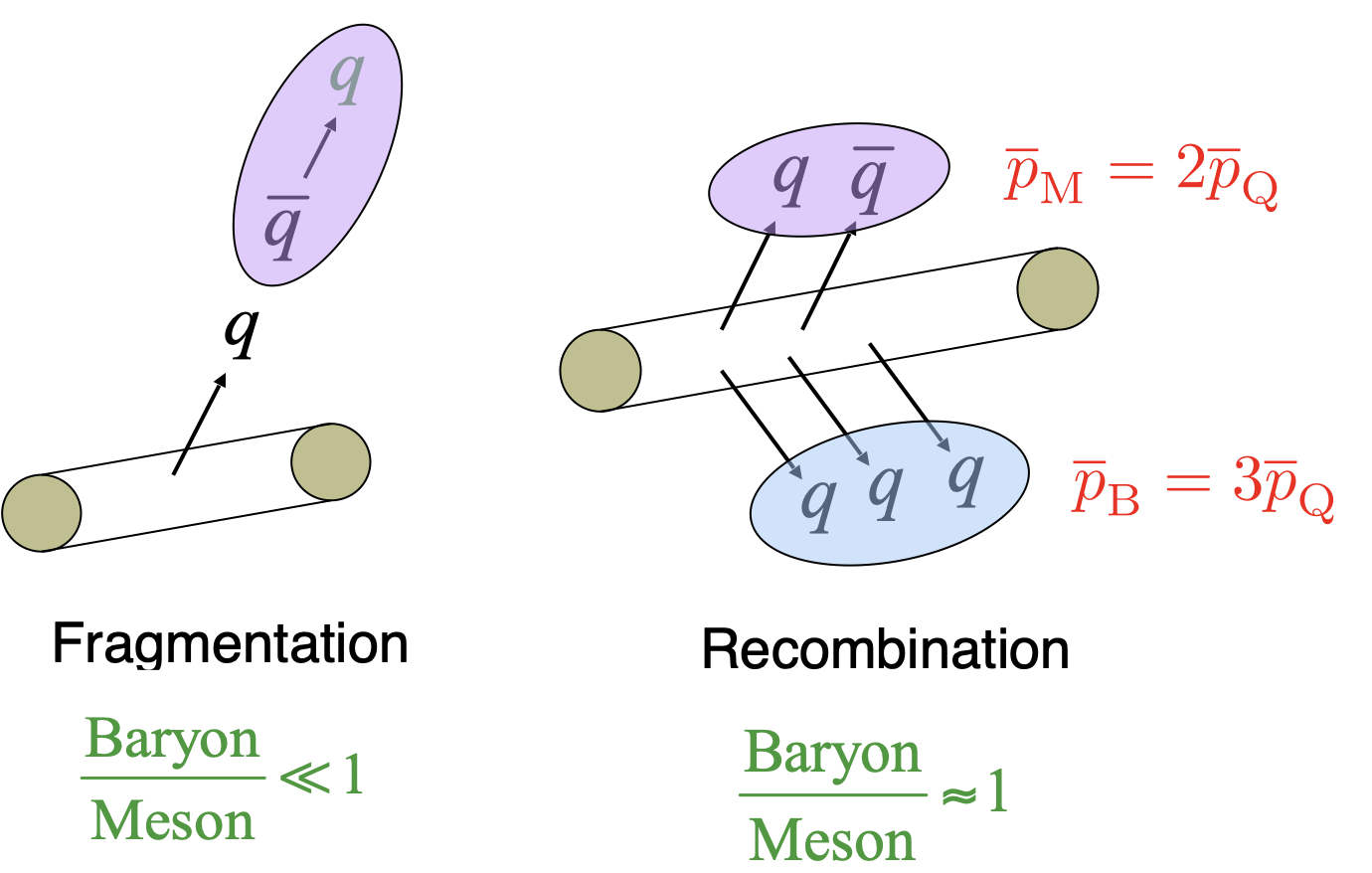}
	\caption{Hadronization mechanisms: Fragmentation (left) and valence quark recombination (right).}
	\label{fig7:hadronization}
\end{figure}

When a QGP disassembles, there are two mechanisms that can lead to the production of hadrons (see Fig.~\ref{fig7:hadronization}). \index{hadronization mechanisms} One is fragmentation, where an energetic quark or anti-quark tries to escape from the fireball and produces another partner to form a meson. \index{fragmentation} It's much less likely to produce a baryon in this manner, and therefore the baryon-to-meson ratio should be much less than unity, which is what is observed in proton-proton collisions. On the other hand, you can have recombination, \index{recombination} \index{coalescence} where a quark from the QGP finds a partner anti-quark inside the QGP, and they escape together. In this case, the meson momentum is approximately twice the average quark momentum. Or three quarks find each other and escape as a baryon, in which case the typical momentum would be three times the quark momentum. Rainer Fries, Chiho Nonaka, Steffen Bass and I managed to show in 2003 on the basis of kinematic arguments that, in this case, the ratio of baryons to mesons is of order unity~\cite{Fries:2003vb,Fries:2003kq}..

\begin{figure}[tb]
	\centering
	\includegraphics[width=0.8\linewidth]{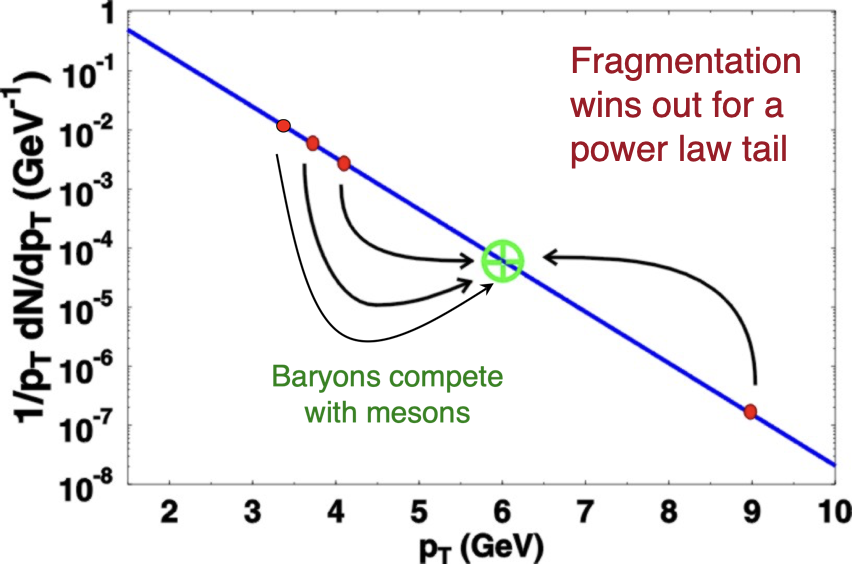}
	\caption{Illustration showing why recombination wins out over fragmentation for an exponential quark spectrum, and why 
	recombination of quark triplets into baryons competes favorably with recombination quark-antiquark pairs into mesons.}
	\label{fig8:reco-vs-frag}
\end{figure}
 
In the kinematic domain where the hadrons escape quickly from the plasma, that is for intermediate transverse momenta in the range of a few GeV/$c$, one can show that for a thermal source the recombination mechanism always dominates over fragmentation ~\cite{Fries:2003rf}. Whereas, for a power law spectrum, which prevails in the high-momentum tail of the spectrum of partons produced in a collision, the fragmentation process always wins out. Furthermore, for an exponential spectrum, like a thermal spectrum, baryons compete favorably with mesons. This is illustrated in Fig.~\ref{fig8:reco-vs-frag}. 

\begin{figure}[th]
	\centering
	\includegraphics[width=0.8\linewidth]{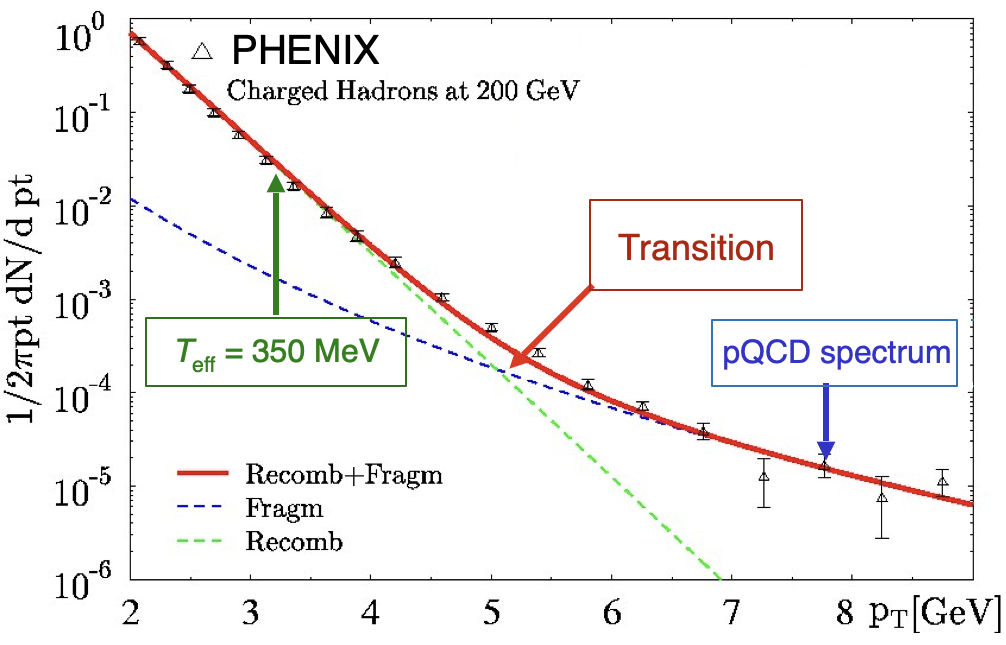}
	\caption{The transverse momentum spectrum of charged hadrons measured by PHENIX in Au+AU collisions at RHIC is explained 
	by the competition between recombination and fragmentation mechanisms (adopted from Ref.~\cite{Fries:2003vb}).}
	\label{fig9:reco+frag}
\end{figure}

Recombination and fragmentation, together, provide a very nice fit to the spectra that were measured the PHENIX experiment \index{PHENIX experiment} in collisions of two gold nuclei. This illustrated in Fig.~\ref{fig9:reco+frag}, where you clearly see the evidence of thermal recombination emission at transverse momenta up to a few GeV/$c$, and then fragmentation taking over and dominating at high transverse momenta~\cite{Fries:2003vb}.

Perhaps the most exciting part of that picture was pointed out by Molnar and Voloshin~\cite{Molnar:2003ff}. If hadrons are formed by recombination of valence quarks, then the flow pattern that is ingrained in the recombining quarks and anti-quarks carries over into the produced hadrons. \index{quark number scaling} In heavy ion collisions one observes a phenomenon called {\em elliptic flow}, reflecting the initial oblong shape of the QGP fireball, that results in an anisotropic pattern of emission with respect to the collision axis. One finds that the pattern for mesons is twice as strong as for the quarks, but at twice the momentum of the individual quarks, because mesons contain a quark and an anti-quark and, therefore, carry on average twice the quark momentum. The factor two would be replaced with a factor three in the case of baryons, and so they suggested, one should simply re-scale the hadron momentum by a factor two or three and the flow pattern by the same factor two or three, to recover the anisotropic flow carried by the quarks.

\begin{figure}[tb]
	\centering
	\includegraphics[width=0.45\linewidth]{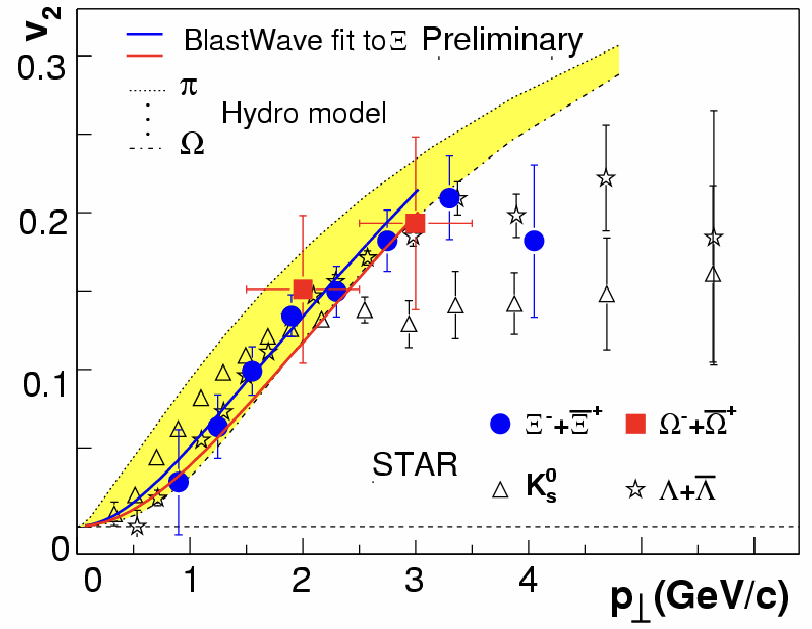}
	\hspace{0.05\linewidth}
	\includegraphics[width=0.45\linewidth]{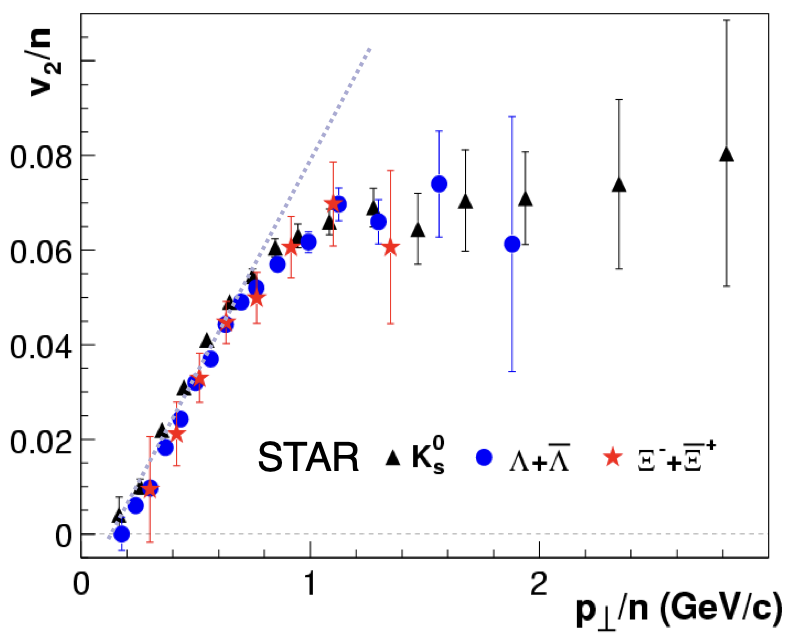}	
	\caption{Left panel: Elliptic transverse flow functions $v_2(p_T)$ measured for different hadrons in Au+Au collisions at RHIC.
	Right panel: Data collapsing into a single curve after scaling with the number $n$ of valence quarks $v_2(p_T/n)/n$ 
	(adopted from Ref.~\cite{Oldenburg:2004qa}).}
	\label{fig10:scaling-RHIC}
\end{figure}

\begin{figure}[tb]
	\centering
	\includegraphics[width=0.45\linewidth]{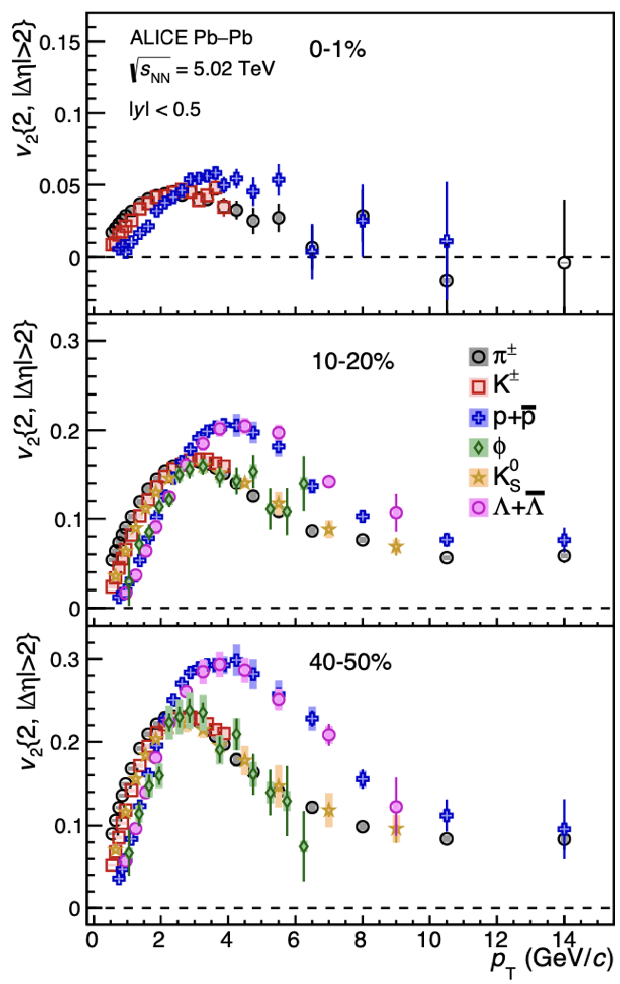}
	\hspace{0.05\linewidth}
	\includegraphics[width=0.45\linewidth]{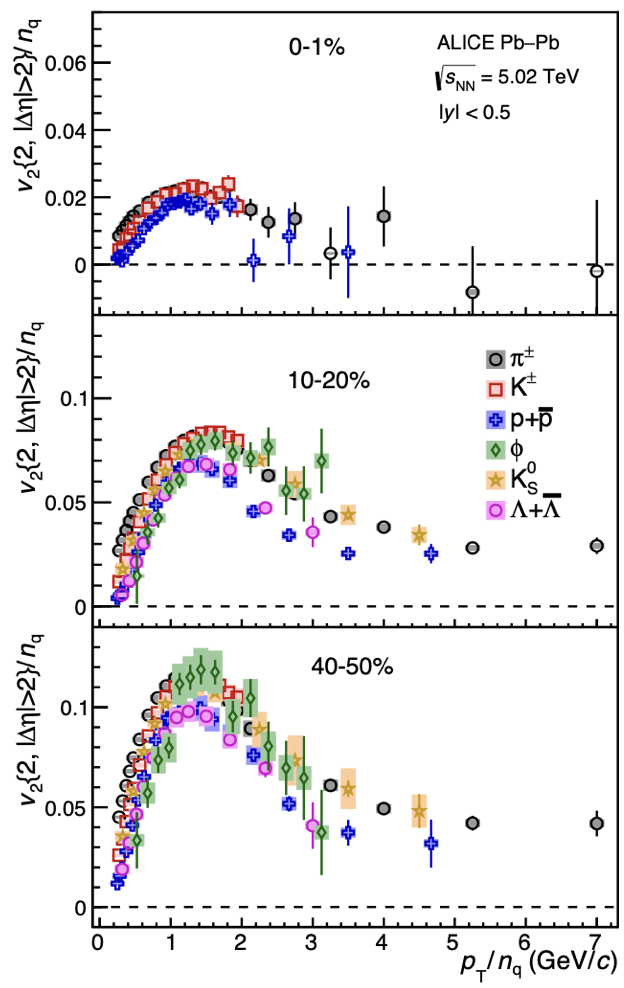}	
	\caption{Valence quark scaling of elliptic flow  observed in Pb+Pb collisions at LHC. 
	Left panel: original data; right panel: re-scaled data (adopted from Ref.~\cite{Acharya:2018zuq}).}
	\label{fig11:scaling-LHC}
\end{figure}

Indeed, one finds that the different curves $v_2(p_T)$ for different hadrons coalesce into a single one as shown in Fig.~\ref{fig10:scaling-RHIC}, in other words the scaling was clearly observed~\cite{Oldenburg:2004qa}. There is much better data now from RHIC, \index{RHIC data} and they have all confirmed the valence quark scaling; this is also true at the LHC. Figure \ref{fig11:scaling-LHC} shows results from ALICE \index{ALICE experiment} for various mesons and baryons in Pb+Pb collisions at LHC~\cite{Acharya:2018zuq}. You see clearly that baryons and mesons have a different flow pattern as a function of transverse momentum, but when you re-scale, you find that the patterns coincide within the experimental uncertainties. In other words, the data is consistent with the interpretation that the collective flow in the QGP is carried by separate unconfined quarks that then imprints itself on the pattern of hadrons at hadronization. \index{deconfinement}

\subsection{Fluctuations}
\label{subsec:fluctuations}

Fluctuations of conserved quantities, \index{fluctuations} \index{conserved quantum numbers} my next topic, are interesting because first of all, they survive final-state interactions as they cannot be locally changed by dynamical processes -- after all, they are conserved -- and they change drastically between a free quark system and a hadronic system~\cite{Asakawa:2000wh,Jeon:2000wg}. This is because hadrons have integer charges and baryon numbers, while quarks carry fractional charges and baryon numbers. Up to now, they have been important, but not central, diagnostic tools (see e.g.~\cite{Borsanyi:2011sw}); this will likely change in the near future when better data from RHIC for the formation of QGP near a possible critical point will become available. \index{critical point} The primary tool for this investigation will be critical fluctuations~\cite{Stephanov:1999zu,Mroczek:2020rpm}. \index{critical fluctuations} As scientists begin to look at the data from the high-statistics RHIC beam energy scan that has just concluded, they will analyze the data for the possible presence of such fluctuations because they can tell us something about the existence of a critical point in the QCD phase diagram.

\section{Diagnostics sensitive to gluons} 
\label{sec:diag-gluons}

\begin{figure}[tb]
	\centering
	\includegraphics[width=\linewidth]{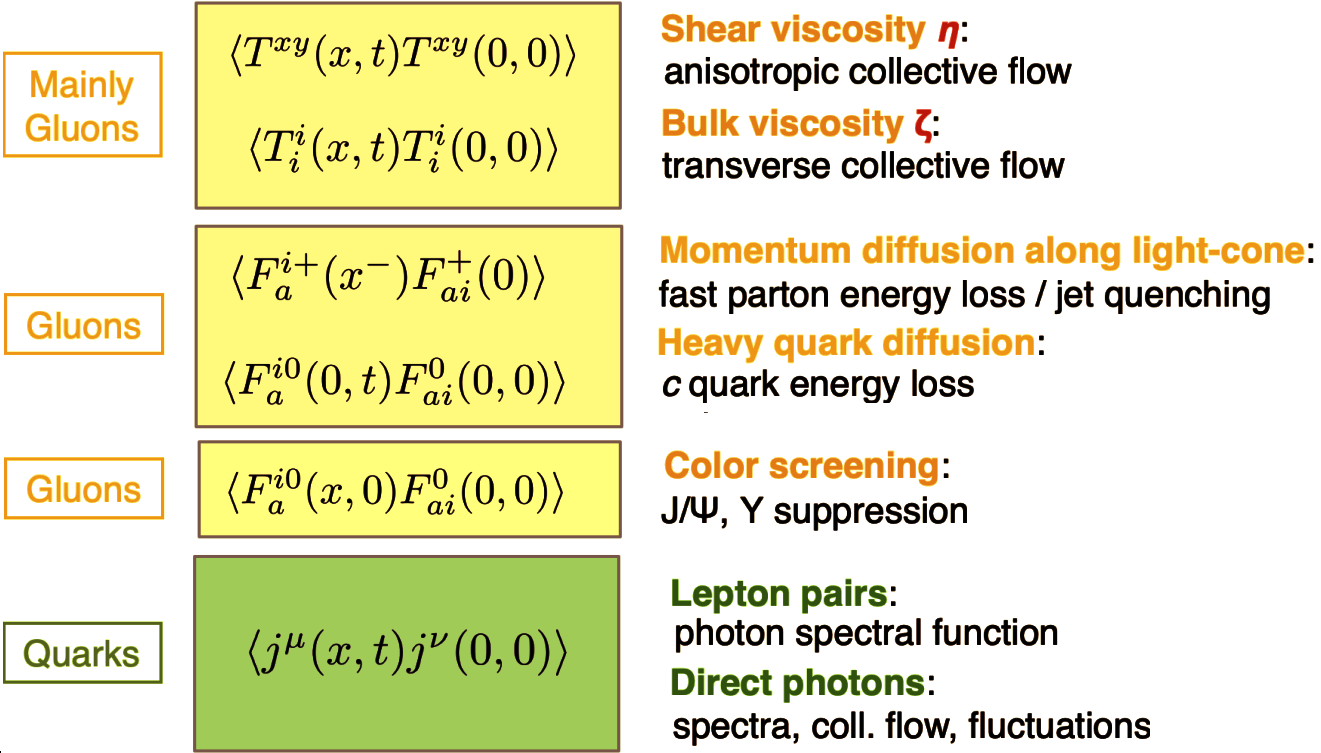}
	\caption{Fundamental properties of the QGP that are accessible to measurements in relativistic heavy ion collisions.}
	\label{fig12:QGP-observables}
\end{figure}

Let me now turn to diagnostics that are sensitive to gluons. \index{gluons} \index{QGP diagnostics} The gluons, obviously, are fundamental components of the QGP. There are certain fundamental properties of the QGP that are accessible to measurements in relativistic heavy ion collisions. Figure \ref{fig12:QGP-observables} list some important examples~\cite{Muller:1994rb}:
\begin{itemize}
\item
Correlations of the components of the energy momentum tensor $T^{\mu\nu}$, which reveal themselves in the shear and bulk viscosity that influence the expansion of the QGP and its flow patterns. 
\item
Correlations of gluons along the light cone, which are directly observable through the energy loss of high-energy partons that produce jets. 
\item 
Heavy quark diffusion measures gluon correlations in time at a given position, if the quark moves slowly.
\item
Instantaneous, spatial correlations are related to the phenomenon called {\em color screening}, where the potential between two heavy quarks is screened within the QGP, which leads to the suppression of heavy-quark bound states. 
\end{itemize}
In addition, there exist correlations of electric currents due to quarks, which can be observed through electron-positron pairs and direct photons.

\subsection{Viscosity}
\label{subsec:viscosity}

The shear viscosity $\eta$ is a measure of the ease of momentum transport within the QGP. Since the mean free path, for which partons in the QGP can move around before scattering, decreases with increased coupling, $\eta$ is a measure of the interaction among quarks and gluons~\cite{Danielewicz:1984ww}. A large value of $\eta$ corresponds to weak coupling, a small value implies strong coupling. In the strong coupling regime one expects that the ratio of shear viscosity to entropy density $\eta/s$, which is a dimensionless number, is of the order of $1/(4\pi)$~\cite{Kovtun:2004de}. \index{viscosity, shear}

The bulk viscosity $\zeta$ represents a measure of the deviation from scale invariance. Scale invariance is broken in QCD by quantum effects; it is predominantly visible in the QGP near the transition region~\cite{Kharzeev:2007wb}. And so, if one can deduce the value of $\zeta/s$ from the data, one learn something about the gluon and the energy momentum tensor in the QGP. \index{viscosity, bulk}

Theory is fine, but one needs experimental observables and a way to extract the values of these quantities from the experimental observables. Over the past several years, the way in which the model--data comparison has been done for relativistic heavy ion collisions has been revolutionized by emulator-aided multi-parameter Bayesian analyses~\cite{Bernhard:2015hxa}. \index{Bayesian analysis} On one hand, we have copious data, both from RHIC and LHC, of spectra, flow velocities, and many other quantities. On the other hand, we have detailed models that describe how the QGP expands and ultimately converts into hadrons and finally breaks up. The combination of these two developments allows us to extract these QGP properties quantitatively and with quantifiable confidence~\cite{Everett:2020xug}.

\begin{figure}[tb]
	\centering
	\includegraphics[width=0.8\linewidth]{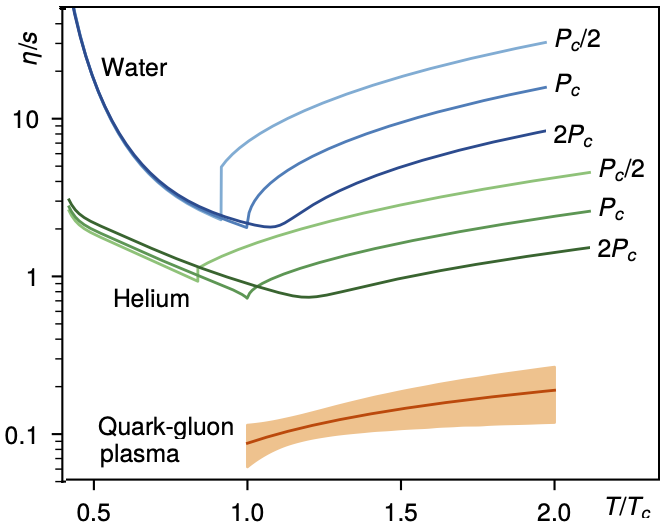}
	\caption{Temperature dependence of the specific shear viscosity $\eta/s$ obtained by a Bayesian model-to-data comparison
	(adopted from Ref.~\cite{Bernhard:2019bmu}).}
	\label{fig13:shearviscosity}
\end{figure}

That was really first demonstrated in great detail in a Ph.D.\ thesis by Jonah Bernhard at Duke, who performed a complete Bayesian sensitivity analysis for a large number of parameters that are related to fundamental properties of the QGP, including the equation of state and the viscosities~\cite{Bernhard:2019bmu}. If we focus on the parameters that are relevant for the shear viscosity they show that, with well controlled uncertainty, the specific shear viscosity in the QGP is indeed much smaller than those in common materials, like helium or water, as shown in Fig.~\ref{fig13:shearviscosity}. The results confirm that QGP is very close to the putative quantum bound $\eta/s = 1/(4\pi)$, thereby confirming the ``perfect fluid'' property of the QGP. \index{prefect fluid}

\begin{figure}[tb]
	\centering
	\includegraphics[width=0.8\linewidth]{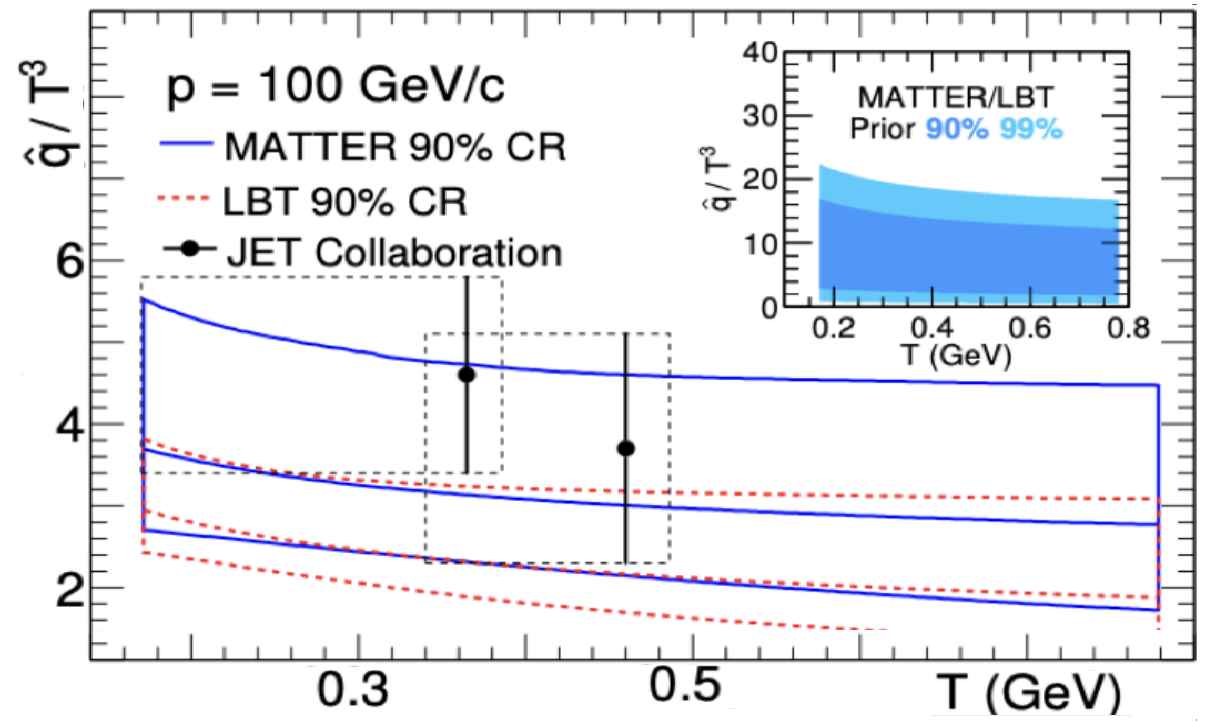}
	\caption{The normalized jet quenching parameter $\hat{q}/T^3$ deduced from RHIC and LHC data on jet suppression
	(adopted from Ref.~\cite{Cao:2021keo}).}
	\label{fig14:qhat}
\end{figure}

We next turn to jet quenching. \index{jet quenching} As I mentioned in conjunction with the Table in Fig.~\ref{fig12:QGP-observables}, the energy loss of fast partons in the QGP depends on the light-cone correlator of glue fields. There is a parameter that one usually calls $\hat{q}$, which is proportional to the radiative energy loss per unit distance of a fast parton propagating through the QGP~\cite{Baier:1996kr}. By performing an analysis of the suppression of energetic hadron production in a heavy ion collision, one can deduce the value of $\hat{q}$. The first determination was done by the so-called JET Collaboration~\cite{Burke:2013yra}, \index{JET collaboration} a  medium sized collaboration of theorists, which has been extended into the JETSCAPE \index{JETSCAPE collaboration} and, more recently, X-SCAPE Collaborations. \index{X-SCAPE collaboration} The participant scientists have determined that the ratio of $\hat{q}/T^3$ is somewhere in the range of four~\cite{Cao:2021keo}, see Fig.~\ref{fig14:qhat}. This implies a quantitative determination of the gluon density in the QGP, as measured by a particle that travels at the speed of light.

\subsection{Color screening}
\label{subsec:screening}

\begin{figure}[tb]
	\centering
	\includegraphics[width=0.21\linewidth]{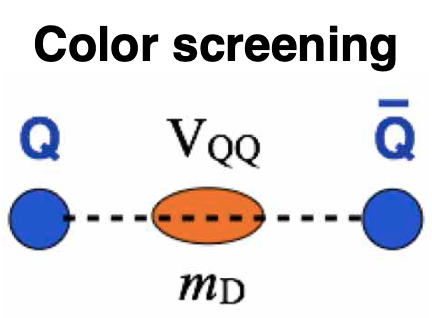}
	\hspace{0.1\linewidth}
	\includegraphics[width=0.28\linewidth]{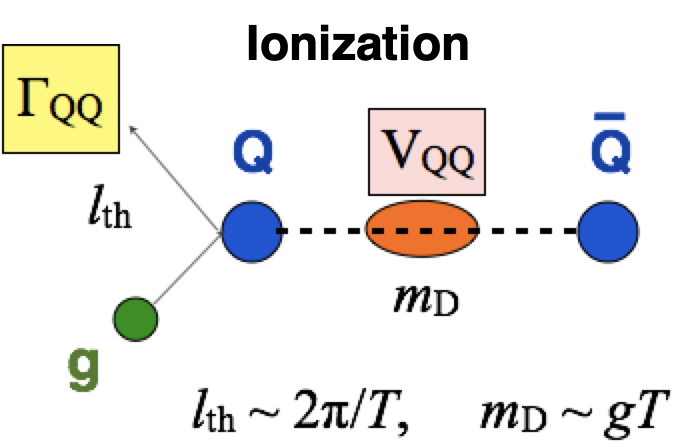}
	\hspace{0.1\linewidth}
	\includegraphics[width=0.2\linewidth]{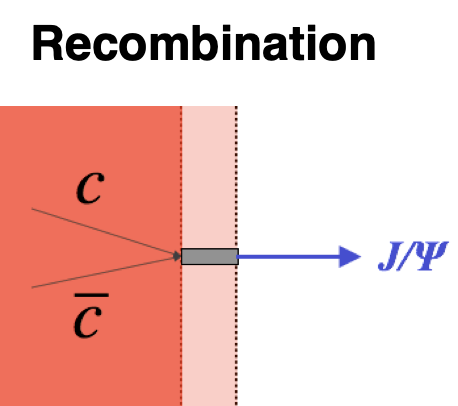}
	\caption{Mechanisms influencing the yield of heavy quarkonia in relativistic heavy ion collisions. Left: Color screening in the QGP
	causes quarkonium states to sequentially ``melt'' if their size exceeds the screening length. Center: Interaction with thermal gluons
	leads to the ionization of heavy quarkonia below the melting point. Right: Heavy quark bound states can be formed at hadronization
	by recombination of a heavy quark and antiquark.}
	\label{fig15:quarkonia}
\end{figure}

On the other hand, if heavy quarks imbedded in the QGP travels slowly, it sees gluon correlations either in the spatial direction, if one looks at the potential that binds them, or in the temporal of direction, if one looks at diffusion phenomena. Correlations in space are responsible for the screening of the color force, quantified by the so-called Debye mass $m_\mathrm{D}$. \index{Debye mass} \index{color screening} This screening leads, as Matsui and Satz predicted~\cite{Matsui:1986dk}, to the dissolution or melting of the bound states of heavy quarks in the QGP.  The left panel of Fig.~\ref{fig15:quarkonia} \index{quarkonium suppression} shows a schematic view of this mechanism: the potential between two heavy quarks, say a  $b$-quark and a $\overline{b}$-quark, is screened on a distance scale given by the inverse the Debye mass $m_\mathrm{D}$. 

When one compares with real data it turns out that this is not the only process that contributes to the ``melting'' of heavy-quark bound states. There are several other processes that are also important: For example, the ionization of heavy-quark bound state under bombardment by thermal gluons (see central panel of Fig.~\ref{fig15:quarkonia}). Or if there is an abundance of heavy quarks, in particular $c$-quarks ($b$-quarks are not as abundantly produced), they can re-form a heavy-quark bound state when they hadronize (see right-most panel of Fig.~\ref{fig15:quarkonia}). In general, there is a competition between ionization and recombination.

\subsection{Jet quenching}
\label{subsec:jetquenching}

\begin{figure}[tb]
	\centering
	\includegraphics[width=0.8\linewidth]{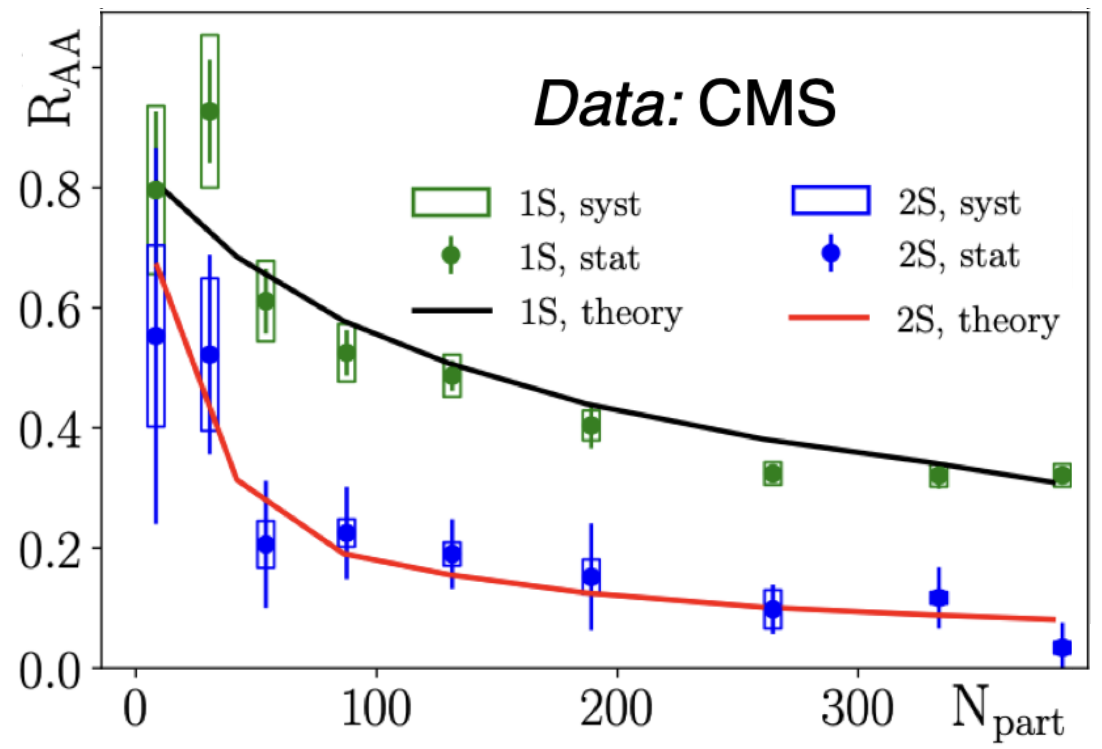}
	\caption{Suppression factor $R_{AA}$ of the $b\overline{b}$ bound states $\Upsilon$(1s) and $\Upsilon${2s} in Pb+Pb collisions
	at LHC relative to their scaled production in p+p collisions. The curves are fits to the CMS data using the open quantum system
	approach (adopted from Ref.~\cite{Yao:2020kqy}).}
	\label{fig16:Upsilon}
\end{figure}

Over the past few years, the transport theory for heavy quarks in the QGP has been formulated rigorously using the various relevant mass scales that are involved. There is a formalism, called {\em open quantum systems}, \index{open quantum system} in which the quarkonium bound state is considered in interaction with the surrounding thermal medium~\cite{Akamatsu:2011se}. This formulation makes use of effective field theories, as appropriate for heavy quarks, and the input in calculation is expressed by a small number of rigorously defined calculable quantities~\cite{Yao:2020kqy,Akamatsu:2020ypb,Brambilla:2020qwo,Yao:2021lus}. Figure~\ref{fig16:Upsilon} shows an example of such a calculation, done last year by Xiaojun Yao, a former Duke graduate student now at MIT, who showed that with reasonable values for these parameters one could nicely explain the suppression of the heavy Upsilon ($b\overline{b}$) states, the 1S state and the 2S state, in a QGP~\cite{Yao:2020kqy,Yao:2020xzw}. \index{Upsilon suppression} In the future we will get much higher statistics data coming from RHIC with the sPHENIX detector and also from the LHC.s and the measurement of ingredients into these calculations at the EIC, namely nuclear parton distribution functions, that will reduce these still sizable uncertainties in the theory data comparison.
 
\section{Summary}
\label{sec:diag-summary}

Let me summarize. We started out with the ``smoking guns'', but the field has progressed from this into what one might call forensic analysis, where the diagnostics of the QGP have moved from the original speculative ideas to a portfolio of rigorous theory-based analyses. This development has been aided by the systematic formulation of observables using effective field theory and transport parameters that connect to well defined properties of the plasma and the data. The Bayesian multi-parameter model data comparison that has become available in the last few years finally permits the extraction of these parameters from the experimental data with quantifiable uncertainties.

So what have we learned? Instead of attempting a full overview of all the insights here, let me mention a few:
\begin{itemize}
\item 
The QGP contains unconfined collectively flowing quarks we know this from the success in many aspects of the quark recombination picture. 
\item
We have multiple and consistent determinations of the gluon composition of QGP, from jet quenching, from the viscosities, and from color screening.
\item
We have quantitatively established from the data that color screening and the near perfect fluidity are properties of the QGP. 
\end{itemize}
There is the expectation that, in the next five years or so, we will get much better data from the RHIC beam energy scan, \index{beam energy scan} from the sPHENIX \index{sPHENIX experiment} run at RHIC, and from the LHC Run 3, with upgraded detectors that will improve the textbook results that I have covered.

\section{Acknowledgments}
\label{sec:diag-ack}

I am deeply honored to receive the 2021 Herman Feshbach prize. Herman was one of the scientists I greatly admired, not only because of his scientific achievements, but also for his sustained work on behalf of scientists behind the Iron Curtain, for his engagement for nuclear disarmament, and his general concern about the equality of opportunity for everybody in the sciences. I sincerely thank all the current and former members of the Duke QCD Theory group. The research summarized here would have been impossible without their major contributions and their collaboration. I am also grateful to U.~Heinz, J.-F.~Paquet, J.~Rafelski, and W.~A.~Zajc for advice during the preparation of the lecture, and C.~M. Grayson for help in editing the transcript. This work was supported by the Office of Science of the U.~S.~Department of Energy under Grant DE-FG02-05ER41367.

\end{document}